%% file: proceedings_1.tex
\documentclass[12pt,longbibliography]{article}
\usepackage{graphicx}
\usepackage{changepage}
\usepackage[symbol]{footmisc}
\usepackage{amsmath} 
\usepackage{amssymb} 
\usepackage{slashed}
\usepackage{inputenc}
\usepackage{color}

\newcommand\pubnumber{CIPANP2018-Fornal}
\newcommand\pubdate{\today}

\def\UCSD{Department of Physics\\
 University of California, San Diego\\
  9500 Gilman Drive, La Jolla, CA 92093, USA}
\def\support{\footnote{\,Research supported in part by the DOE Grant No.~${\rm DE}$-${\rm SC0009919}$.\label{xx}}}

\def\Title#1{\begin{center} {\Large #1 } \end{center}}
\def\Author#1{\begin{center}{ \sc #1} \end{center}}
\def\Address#1{\begin{center}{ \it #1} \end{center}}

\newcommand\pubblock{\rightline{\begin{tabular}{l} \pubnumber\\
         \pubdate  \end{tabular}}}
\newenvironment{Abstract}{\begin{quotation}  }{\end{quotation}}
\newenvironment{Presented}{\begin{quotation} \begin{center} 
             PRESENTED AT\end{center}\bigskip 
      \begin{center}\begin{large}}{\end{large}\end{center} \end{quotation}}

\input econfmacros.tex

\def\bea{\begin{eqnarray}}
\def\eea{\end{eqnarray}}
\def\bean{\begin{equation*}}
\def\eean{\end{equation*}}

\begin{document}
\begin{titlepage}

\pubblock
\renewcommand{\thefootnote}{\fnsymbol{footnote}} 

\vfill
\Title{Neutron Lifetime Discrepancy as a Sign of a Dark Sector?}
\vfill
\Author{ Bartosz~Fornal\support\footnote{\,Speaker; talk title: \emph{Dark Matter Interpretation of the Neutron Decay Anomaly};  \\ \indent \ \hspace{2mm} based on: B. Fornal and B. Grinstein, Phys.\,Rev.\,Lett.\,120,\,191801\,(2018) \cite{Fornal:2018eol}}  and Benjam\'{i}n~Grinstein$^{\ref{xx}}$}
\Address{\UCSD}
\vfill
\begin{adjustwidth}{-1mm}{-1mm}
\begin{Abstract}
We summarize our recent proposal of explaining the discrepancy between the bottle and beam measurements of the neutron lifetime through the existence of a dark sector, which the neutron can decay to with a branching fraction 1\%. We show that viable particle physics models for such neutron dark decays can be constructed and we briefly comment on recent developments in this area. 
\end{Abstract}
\end{adjustwidth}
\vfill
\begin{adjustwidth}{-6mm}{-6mm}
\begin{Presented}
13$^{\rm th}$ Conference on the Intersections of Particle and Nuclear Physics\\ \vspace{3mm}
Palm Springs, CA, USA, May 29 -- June 3, 2018
\end{Presented} 
\vfill
\end{adjustwidth}
\end{titlepage}
\def\thefootnote{\fnsymbol{footnote}}
\setcounter{footnote}{0}

\section{Neutron Lifetime}

In the Standard Model (SM), the dominant neutron decay channel is beta decay,
\bea
n \rightarrow p + e^- \!+ \bar\nu_e \ ,
\eea
along with  radiative corrections involving one or more photons in the final state.
The theoretical prediction for the neutron lifetime  is \cite{Marciano:2005eca}
\bea\label{life}
\tau_n^{\rm SM} = \frac{4908.7(1.9)\,{\rm s}}{|V_{ud}|^2(1+3g_A^2)} \ ,
\eea
where $g_A$ is the ratio of the axial-vector to vector coupling. By using  the average $V_{ud}$ and $g_A$ values extracted from experiments and reported by the Particle Data Group (PDG) \cite{Tanabashi:2018oca}, one gets 
$875.3 \ {\rm s}< \tau_n < 891.2 \ {\rm s}$ within $3\,\sigma$. In turn, a recent lattice calculation of $g_A$ \cite{Chang:2018uxx,Berkowitz:2018gqe} yields $\tau_n = 885 \pm 15 \ {\rm s}$.

On the experimental side, there are two types of  neutron lifetime measurements: the bottle method and the beam method. 

In bottle experiments, neutrons are trapped in a container and their number ($N_n$) is determined as a function of time. The decaying exponential  is fit to the data, 
\bea
N_n(t) = N_n(0) \,e^{-t/\tau^{\rm bottle}_n}
\eea
and the neutron lifetime is extracted. Those types of experiments measure the total neutron lifetime.
The PDG average based on five bottle experiments \cite{Mampe,Serebrov:2004zf,Pichlmaier:2010zz,Steyerl:2012zz,Arzumanov:2015tea} is
\bea
\tau_n^{\rm bottle}  = 879.6 \pm 0.6  \ {\rm s} \ ,
\eea
with recent measurements \cite{Serebrov:2017bzo,Pattie:2017vsj} lying within $2\, \sigma$ of this average.

The beam experiments, on the other hand, arrive at the neutron decay rate  by counting the  resulting protons ($N_p$). Determining also the number of neutrons in the beam ($N_n$)  which those protons came from, the neutron lifetime is obtained via
\bea
\frac{1}{\tau^{\rm beam}_n} = -\frac{1}{N_n}\frac{d N_p}{dt} = -\frac{{\rm Br}(n\rightarrow p + {\rm anything})}{N_n}\,\frac{d N_n}{dt}\ .
\eea
The SM predicts ${\rm Br}(n\rightarrow p \,+\, {\rm anything}) = 1$, so without new physics the two lifetimes should be identical, $\tau^{\rm beam}_n= \tau^{\rm bottle}_n$. However, if there exists some other neutron decay channel(s) with no protons in the final state, then ${\rm Br}(n\rightarrow p + {\rm anything}) < 1$ and
\bea\label{ineq}
\tau^{\rm beam}_n =  \frac{\tau_n}{{\rm Br}(n\rightarrow p + {\rm anything})} > \tau^{\rm bottle}_n \ . 
\eea
The PDG average based on two beam experiments \cite{Byrne:1996zz,Yue:2013qrc} (see also Ref.\,\cite{Nico:2004ie}) is 
\bea
\tau_n^{\rm beam}  = 888.0 \pm 2.0  \ {\rm s} \ ,
\eea
which is in $4.0 \, \sigma$ tension with the bottle result and suggests that Eq.~(\ref{ineq}) holds. 

\newpage

This discrepancy might be a result of some  systematic uncertainties that were not accounted for. However, it may well be an effect of  new physics. We explore the latter possibility, for which the bottle and beam results imply
\bea
{{\rm Br}(n\rightarrow p + {\rm anything})} \approx 0.99 \ ,
\eea
with the remaining 1\% coming from \emph{neutron dark decays} involving, but not limited to, particles beyond the SM in the final state.

 \section{Neutron Dark Decay}
 \vspace{-0.4mm}
 
 Let us denote  the final state of neutron dark decay by $f$ and the sum of the final state particle masses by $M_f$. Obviously, the upper bound is the neutron mass, $M_f < m_n$. The lower bound on $M_f$ arises from nuclear stability. The SNO  \cite{Ahmed:2003sy} and KamLAND  \cite{Araki:2005jt} experiments looked for signatures of a neutron disappearing inside $^{16}{\rm O}$ and $^{12}{\rm C}$ nuclei. The scenarios considered involved  the daughter nucleus in an excited state, leading to subsequent de-excitation with the emission of secondary particles, e.g. gamma rays. The bound on the neutron lifetime of $\tau_{n\to {\rm invisible}}> 5.8\times 10^{29}$ years was established, a value adopted  by the PDG as the constraint on neutron invisible decays. However, if $M_f$ is close enough to $m_n$, i.e., $m_n - S_n< M_f < m_n$, where $S_n$ is the neutron separation energy in a given nucleus, the neutron dark decay would remain  allowed, while the corresponding nuclear decay would be kinematically forbidden. 
 
 The tightest bound of this type comes from the requirement of $^9{\rm Be}$ stability. The $^9{\rm Be}$ nucleus has a neutron separation energy of $S_n(^9{\rm Be}) = 1.664 \ {\rm MeV}$. If kinematically allowed, a neutron dark decay would lead to $^9{\rm Be} \to \!\,^8{\rm Be} + f$. However, this nuclear decay is forbidden if $M_f > m_n - S_n(^9{\rm Be})$, leading ultimately to the requirement
\bea\label{con}
937.900 \ {\rm MeV} <  M_f <  939.565 \ {\rm MeV} \ ,
\eea
which also assures proton stability. 

The nonzero mass window  in Eq.~(\ref{con}) opens the door to a whole rage of possibilities for the neutron dark decay final states:  dark particle + photon, two dark particles, dark particle + $e^+e^-$ pair, etc. Below we focus on the first two cases.

\subsection{${\boldsymbol {\rm Neutron \to dark \ particle +photon}}$}
 In this minimal case, the neutron dark decay $n \to \chi\,\gamma$ final state consists of a dark particle $\chi$  with a mass in the range
 \bea
 937.900 \ {\rm MeV} <  m_\chi <  939.565 \ {\rm MeV} \, ,
 \eea
and a monochromatic photon with energy anywhere between
 \bea\label{phE}
0 < E_\gamma < 1.664 \ {\rm MeV} \, ,
\eea
where the lower bound corresponds to $m_\chi$ being close to $m_n$.

\newpage

 If $\chi$ is a dark matter particle, it then has to be stable. In particular, it cannot decay to a proton and an electron, thus $m_\chi < m_p+m_e$. This leads to a narrower allowed energy range for the photon, $ 0.782 \ {\rm MeV} <E_\gamma < 1.664 \ {\rm MeV} $.
 \vspace{2mm}
 
 An example of an effective theory describing the decay $n \to \chi\,\gamma$ is given by 
 \bea\label{lageff11}
\mathcal{L}^{\rm eff}_1 = \bar{n}\,\big(i\slashed\partial-m_n +\tfrac{g_ne}{2 m_n}\sigma^{\mu\nu}F_{\mu\nu}\big) \,n
+   \bar{\chi}\left(i\slashed\partial-m_\chi\right) \chi + \varepsilon \left(\bar{n}\chi + \bar{\chi}n\right) \ ,
\eea
where $g_n$ is the neutron $g$-factor and $\varepsilon$ is a model-dependent mixing parameter with dimension of mass. This Lagrangian yields the neutron dark decay rate 
\bea\label{eff1}
\Delta\Gamma_{n\rightarrow \chi\gamma} = \frac{g_n^2e^2}{8\pi}\bigg(1-\frac{m_\chi^2}{m_n^2}\bigg)^3  \frac{m_n\,\varepsilon^2}{(m_n-m_\chi)^2} \  ,
\eea
which needs to be $\approx 1\%$ of the total neutron decay rate to explain the discrepancy between bottle and beam experiments. A viable particle physics model for the dark decay $n \to \chi\,\gamma$ is provided by Model 1 in Sec.~\ref{mod1}.
\vspace{1mm}

\subsection{${\boldsymbol {{\rm Neutron \to two \ dark \ particles} }}$ }
The second, truly dark scenario involves  a dark fermion $\chi$ and a dark scalar $\phi$ in the final state, interacting with the neutron via an intermediate fermion $\tilde\chi$. The neutron dark decay is  $\,n\to \tilde\chi^* \to \chi\,\phi\,$ and the condition in Eq.\,(\ref{con}) takes the form
\bea
937.900 \ {\rm MeV} < m_\chi + m_\phi <  939.565 \ {\rm MeV} \ .
\eea
In this case $\chi$ does not need to have a mass close to the neutron mass. In particular, the masses of both $\chi$ and $\phi$ could be $\sim 470 \ {\rm MeV}$. However, the additional requirement
\bea
m_{\tilde\chi} > 937.9 \ {\rm MeV} 
\eea
is necessary to prevent nuclear decay with $\tilde\chi$ in the final state, e.g.
$^9{\rm Be} \to \!\,^8{\rm Be} + \tilde\chi$.
Both particles $\chi$ and $\phi$ may be stable if their masses satisfy $|m_\chi - m_\phi| < m_p + m_e$.
\vspace{1mm}

An effective description of $n\to \chi\,\phi$ is provided by the Lagrangian
\bea\label{efflag2}
\mathcal{L}^{\rm eff}_{2} \, = \,  \mathcal{L}^{\rm eff}_{1}(\chi \rightarrow \tilde\chi) +  \left(\lambda_\phi \,\bar{\tilde{\chi}}\, \chi\, \phi + {\rm h.c.}\right)
+  \bar{\chi}\left(i\slashed\partial-m_{\chi}\right) \chi + \partial_\mu \phi^* \partial^\mu \phi - m_\phi^2 |\phi|^2  , \ \ \ 
\eea
which yields the neutron dark decay rate
\bea\label{lageff3}
\Delta\Gamma_{n\rightarrow \chi\phi} = \frac{|\lambda_\phi|^2}{16\pi}\sqrt{f(x, y)}\, \frac{m_n\,\varepsilon^2}{(m_n-m_{{\tilde\chi}})^2} \ .
\eea
In Eq.\,(\ref{lageff3}), the function $f(x, y) =[(1-x)^2-y^2] \, [(1+x)^2-y^2]^3$, where $x=m_\chi/m_n$ and $y=m_\phi/m_n$. 

If $m_{\tilde\chi} > m_n$, then   $n\to \chi\,\phi$ is the only available neutron dark decay channel and the expression in Eq.\,(\ref{lageff3}) should give $1\%$ of the total neutron decay rate in order to account for the experimental discrepancy. If, on the other hand, $m_{\tilde\chi} < m_n$, then a second decay channel, $n\to \tilde\chi\,\gamma$, is also present and the ratio of the two decay rates is 
\bea
\frac{\Delta\Gamma_{n\rightarrow \tilde\chi\gamma}}{\Delta\Gamma_{n\rightarrow \chi\phi}} = \frac{2g_n^2e^2}{|\lambda_\phi|^2} \frac{(1-\tilde{x}^2)^3 }{\sqrt{f(x, y)}}\ ,
\eea
where $\tilde{x} = m_{\tilde\chi}/m_n$. To explain the neutron lifetime discrepancy the two decay rates,
$\Delta\Gamma_{n\rightarrow \tilde\chi\gamma}$  and $\Delta\Gamma_{n\rightarrow \chi\phi}$,
have to add up to $1\%$ of the total neutron decay width. A  particle physics  realization of the scenario $\,n\to \chi\,\phi\,$ is provided by Model 2 in Sec.~\ref{mod2}.

 \section{Particle Physics Models}\label{sec3}
 Let us first note that  our proposal of neutron dark decay is phenomenological in its nature and the simplest particle physics models provided below serve only as an illustration of selected scenarios, with more complex dark sectors left to be explored.

 \subsection{Model 1}\label{mod1}
 
 \vspace{-4mm}
  \begin{figure}[h!]
\centering
\includegraphics[height=1.5in]{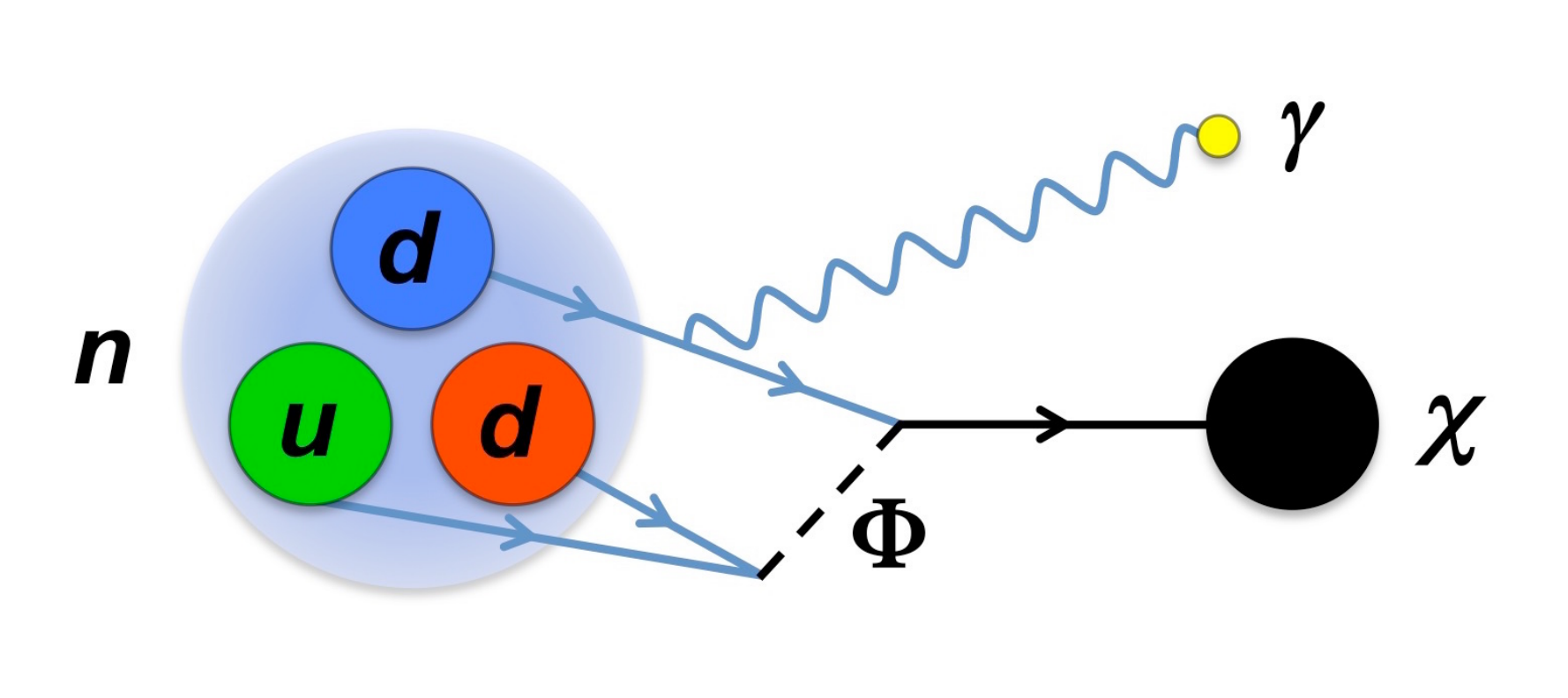}
\vspace{-4mm}
\caption{\small{Neutron dark decay $n \to \chi \, \gamma$ in Model 1.}}
\label{fig:1}
\end{figure}

 The minimal particle physics model for the neutron dark decay involves only two new particles: a Dirac fermion $\chi$ and a colored scalar $\Phi$. We choose $\Phi$ to be a color triplet with hypercharge $-1/3$. 
 The Lagrangian of the model is
 \bea\label{L1}
\mathcal{L}_{1} &\!\!=\!\!&   \Big[ \, \lambda_q \,\epsilon^{ijk}\, \overline{u^c_L}_{i}\, d_{Rj} \Phi_k + \lambda_\chi\Phi^{*i}\bar\chi \,d_{Ri}  + {\rm h.c.}\Big] - M_\Phi^2 |\Phi|^2 - m_\chi \,\bar\chi\,\chi  \ , \ \ \ \ \ \ \ \ 
\eea
where $u^c_L$ is the charge conjugate of $u_R$.
A schematic diagram for the neutron dark decay $n\to \chi\,\gamma$ is shown in Fig.~\ref{fig:1}.

By matching the Lagrangian in Eq.~(\ref{L1}) to the effective Lagranian in Eq.~(\ref{lageff11}), one arrives at the neutron dark decay rate given by Eq.\,(\ref{eff1}) with $\varepsilon = {\beta\,\lambda_q\lambda_\chi }/{M_{\Phi}^2} $, where the matching coefficient $\beta$ is defined via $\langle 0| \epsilon^{ijk} (\overline{u^c_L}_{i} d_{Rj}) d_{Rk}^\rho |n\rangle = \beta \, ({1+\gamma_5})^{\rho}_{\, \sigma}  u^\sigma/2 $ and is calculated from the lattice. 

To reproduce the branching fraction $\Delta\Gamma_{n\rightarrow \chi\gamma} \approx 0.01 / \tau_n$ required to explain the neutron lifetime discrepancy, and taking for example $m_\chi = 937.9 \ {\rm MeV}$, the remaining parameters of the model have to satisfy 
\bea\label{cco}
\frac{M_\Phi}{\sqrt{|\lambda_q\lambda_\chi|}} \approx 400 \ {\rm TeV} \ ,
\eea
which shows that $\Phi$ can easily avoid all collider constraints.
Having a model with $\chi$ being a Dirac particle rather than Majorana makes it free from neutron-antineutron oscillation \cite{Abe:2011ky} and dinucleon  decay \cite{Gustafson:2015qyo} constraints, which would otherwise exclude the model. Note that assigning $B_\chi=1$ and $B_\Phi = -2/3$, Model 1 conserves baryon number.

\subsection{Model 2}\label{mod2}

\vspace{-4mm} 
\begin{figure}[h!]
\centering
\includegraphics[height=1.7in]{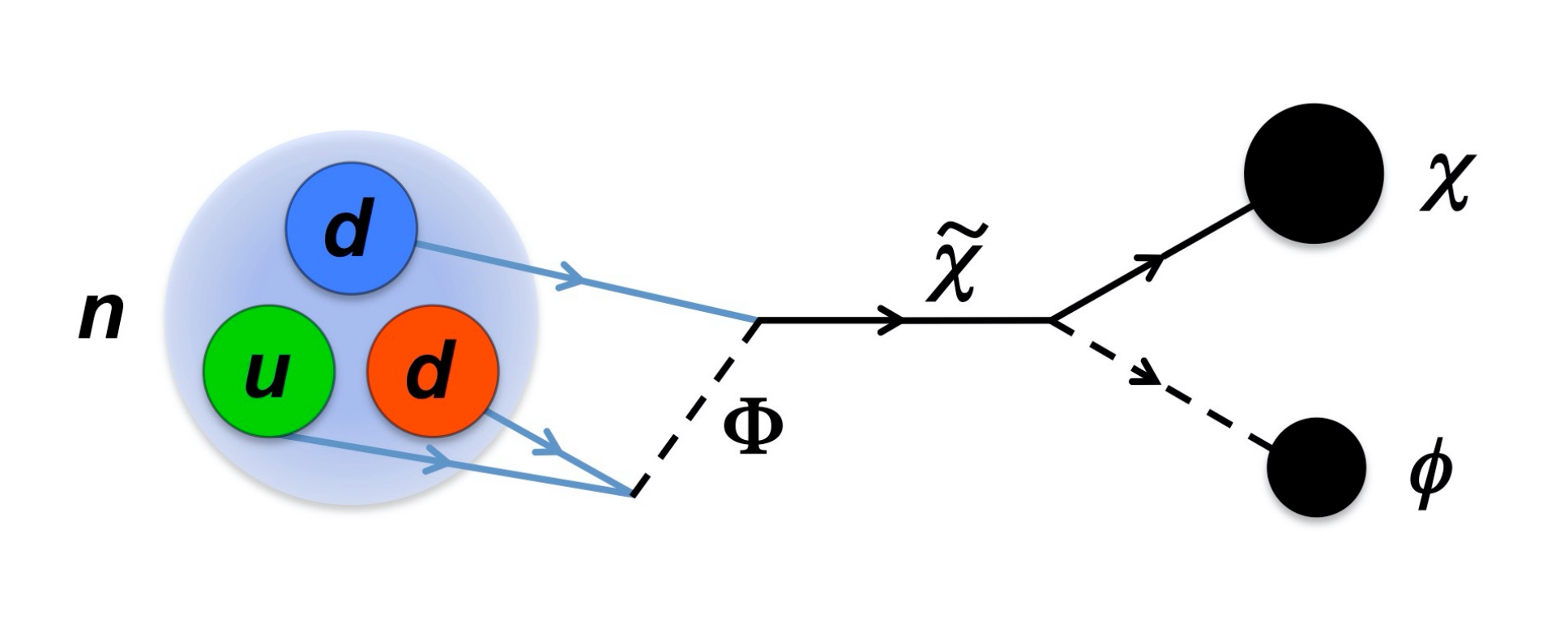}
\vspace{-4mm}
\caption{\small{Neutron dark decay $n \to \chi \, \phi$ in Model 2.}}
\label{fig:2}
\end{figure}

A model with a pure dark neutron decay channel involves four new particles: a Dirac fermion $\chi$ and a scalar $\phi$ in the final state, an intermediate Dirac fermion $\tilde\chi$ and the colored scalar $\Phi$ introduced before (see Fig.~\ref{fig:2}).  The Lagrangian  is similar to the previous one, with  an additional interaction between $\chi$, $\tilde{\chi}$ and $\phi$,
\bea\label{lag245}
\mathcal{L}_{2}  \,=\, \mathcal{L}_{1}(\chi \rightarrow \tilde\chi)   +  ( \lambda_\phi  \,\bar{\tilde\chi}\, \chi \,\phi  + {\rm h.c.})  -  m_\phi^2 |\phi|^2  -  m_\chi \,\bar\chi\,\chi   \ . 
\eea
Assigning $B_{\tilde\chi} = B_\phi=1$ and $B_\chi=0$, baryon number is again conserved. Matching the Lagrangian in Eq.\,(\ref{lag245}) to the effective theory described by the Lagrangian in Eq.\,(\ref{efflag2}), adopting $m_\chi  = 937.9 \ {\rm MeV}$, $m_\phi \approx 0$ and $m_{\tilde\chi} = 2\,m_n$, the required condition on the remaining parameters is
\bea\label{cco2}
\frac{M_\Phi}{\sqrt{|\lambda_q\lambda_{\tilde\chi} \lambda_\phi|}} \approx 300 \ {\rm TeV} \ ,
\eea
again consistent with collider constraints.
As discussed above, if $m_{\tilde\chi}>m_n$ this pure dark decay channel is the only option and its branching fraction should be 1\%.  For $m_{\tilde\chi}< m_n$ the decay $n \to \tilde\chi \,\gamma$  also becomes available, and there is a competition between the two channels, with their rates adding up to 1\%.

 \section{Recent Developments}
 
 Our neutron dark decay proposal inspired several avenues of investigation, both on the theoretical  and the  experimental side. We discuss some of the follow-up papers below.

 \subsection{Theory and model building}
 \vspace{2mm}

  \ \  $\bullet$\  {\bf  \ Neutron stars} \\[5pt]
 In Refs.\,\cite{McKeen:2018xwc,Baym:2018ljz,Motta:2018rxp} the impact of neutron dark decays on the stability of neutron stars was considered. 
 It was shown that for a final state dark particle with  strong repulsive self-interactions, neutron dark decays can be sufficiently blocked such that the observed neutron star masses (up to two solar masses) are allowed. This can be achieved, for instance, by adding to our representative models discussed in Sec.~\ref{sec3} a dark vector boson interacting with the dark particles. Such a strongly interacting dark sector fits well into the paradigm of self-interacting dark matter  (SIDM), 
 which was proposed almost twenty years ago to cure the core-cusp problem and the missing satellite problem of the $\Lambda {\rm CDM}$ cosmological model \cite{Spergel:1999mh}. \\
 
 \noindent
\ \   $\bullet$\  {\bf  \ Models with SIDM} \\[5pt]
 A minimal model of this type, consistent with all astrophysical constraints, was constructed in Ref.\,\cite{Cline:2018ami}, where the neutron dark decay $n \to \chi\, A'$ was considered. The dark photon $A'$ provides the repulsion between the dark particles needed to stay in agreement with the observed neutron star masses. The dark particle $\chi$  can make up  a fraction of the dark matter in the universe. \\[7pt]
 \noindent
A model in which $\chi$ can account for all of the dark matter in the universe was proposed in Ref.\,\cite{Karananas:2018goc}. The resulting neutron dark decay channel is $\,n \to \chi \, \phi\,$, just like in our Model 2. In addition, a vector particle was introduced to induce self-interactions of  $\chi$ and satisfy neutron star constraints. Assuming non-thermal dark matter production in the early universe, this model is consistent with astrophysical constraints and solves the small-scale structure problems of the $\Lambda {\rm CDM}$ paradigm. \\

\noindent
\ \     $\bullet$\  {\bf  \ Hadron dark decays} \\[5pt] 
In Ref.\,\cite{Barducci:2018rlx}  the idea of neutron dark decay was generalized to other neutral hadrons, in particular $K_L^0$ and $B^0$. A representative model for the dark sector was constructed, in which the neutral mesons decay to two dark fermions, whereas the neutron decays to three dark fermions. 
Neutron dark decays in neutron stars are prevented by Pauli blocking, similarly to the SM neutron beta decays.\\

\noindent
\ \     $\bullet$\  {\bf  \ Baryogenesis} \\[5pt] 
It was argued in Ref.\,\cite{Bringmann:2018sbs} that the SIDM model with the neutron dark decay channel $n \to \chi\, A'$ can lead to low-scale baryogenesis. An independent model analogous to our Model 2, in which $\tilde\chi$ couples to additional quark flavors and $\chi$ is a Majorana particle \cite{Elor:2018twp}, was shown to also provide a new mechanism for low-scale baryogenesis. 
\\

\noindent
\ \   $\bullet$\  {\bf  \ Related ideas} \\[3pt]
   It was pointed out in Ref.\,\cite{Czarnecki:2018okw} that the value of $g_A$ in Eq.\,(\ref{life}) obtained from averaging the experimental data taken after the year 2002 favors the bottle result for the neutron lifetime.
Motivated by this, models have been constructed in which it is the bottle measurement that should agree with the SM prediction for the neutron lifetime and not, like in our proposal, the beam result. The explanation of different outcomes of the two types of experiments was done either by postulating neutron-mirror neutron oscillations that are enhanced in the presence of a magnetic field (thus affect only the beam measurement) \cite{Berezhiani:2018eds} or by postulating a large Fierz interference term \cite{Ivanov:2018vit}.
\vspace{4mm}

\subsection{Experimental searches}
\vspace{2mm}

\renewcommand{\thefootnote}{$\mathsection$} 

 \ \  $\bullet$\  \ {${\boldsymbol {\rm Neutron \to dark \ matter +photon}}$} \\[5pt]
Within the first month after our results were announced, a dedicated experimental search for the monochromatic photon from the neutron dark decay $n \to \chi\,\gamma$ was performed \cite{Tang:2018eln}. This search challenged the scenario with ${\rm Br}(n\to \chi\,\gamma) \approx 1 \%$ for a photon of energy in the range $0.782 \ {\rm MeV} < E_\gamma < 1.664 \ {\rm MeV}$ with significance $2.2 \, \sigma$. The case $E_\gamma < 0.782 \ {\rm MeV}$ was not explored. \\

\noindent
\ \ $\bullet$\  \ {${\boldsymbol {{\rm Neutron \to dark \ particle} +{e^+e^-}}}$} \\[5pt]
A dedicated experimental search for the neutron dark decay channel $n\to \chi\,e^+e^-$ \cite{Sun:2018yaw} excluded ${\rm Br}(n\to \chi\,e^+e^-) \approx 1 \%$ for $e^+e^-$ pairs with energies $E_{e^+e^-} \gtrsim 2\,m_e + 100 \ {\rm keV}$ at a very high confidence level.\\

\noindent
\ \ $\bullet$\  \ {\bf Nuclear dark decays} \\[5pt]
 In our original work \cite{Fornal:2018eol} we suggested looking for neutron dark decay-induced decays of unstable nuclei. Such nuclear decays would be possible since some unstable nuclei, e.g. $^{11}{\rm Li}$, $^{11}{\rm Be}$, $^{15}{\rm C}$ and  $^{17}{\rm C}$, have a neutron separation energy $S_n$ smaller than that in $^9{\rm Be}$. Those types of nuclei would undergo dark decays if the dark particle mass happened to be in the range
\bea
937.9 \ {\rm MeV} < m_\chi < m_n - S_n \ .
\eea

We proposed looking for nuclear dark decay signatures using $^{11}{\rm Li}$. However, it was pointed out in Ref.\,\cite{Pfutzner:2018ieu} that $^{11}{\rm Be}$ with $S_n(^{11}{\rm Be}) \approx 0.5 \ {\rm MeV}$ is a better candidate from an experimental perspective. The authors also suggested that the unexpectedly high count of $^{10}{\rm Be}$ in $^{11}{\rm Be}$ decays reported in Ref.\,\cite{Riisager:2014gia} and initially attributed to an enhanced $\beta p$ channel due to an unknown resonance, may in fact be  a sign of the neutron dark decay like  in our Model 2, resulting in the nuclear decay channel\,\footnote{According to Ref.\,\cite{Ejiri:2018dun}, the decay channel $^{11}{\rm Be} \to \!\,^{10}{\rm Be} + \tilde\chi^* \to  \!\,^{10}{\rm Be} + \chi + \phi\,$ is consistent with the observed SM decay rates of $\,^{11}{\rm Be}$ as long as $m_{\tilde\chi} > m_n - S_n(^{11}{\rm Be})$. This condition is well satisfied in the SIDM model of Ref.\,\cite{Karananas:2018goc}, where $m_{\tilde\chi} \approx 800 \ {\rm GeV}$ was adopted.}
\bea
^{11}{\rm Be} \to  \!\,^{10}{\rm Be} + \chi + \phi\ . 
\eea
An experiment determining whether the final state of $^{11}{\rm Be} \!\to\! \!\,^{10}{\rm Be}$ decays contains protons or not has very recently been performed at the CERN-ISOLDE facility \ \ \ \ \ \ \ \ \cite{Pfutzner:2018ieu,ISOLDE}. The results have not yet been published.
\vspace{2mm}

\section{Final Thoughts}

If  the ongoing beam neutron lifetime measurements at NIST (National Institute of Standards and Technology) \cite{NIST2009,NIST} and J-PARC (Japan Proton Accelerator
Research Complex) \cite{Nagakura:2017xmv,Japan} yield results which are still in disagreement with the bottle average, perhaps the best way to proceed would be to include a proton  detection system in the bottle experiment.
With enough precision, such an experiment would be capable of measuring the branching fraction of neutron beta decays independently from the beam experiment. This would directly test the premise that the discrepancy between bottle and beam results is due to neutron decays that do not produce a proton, and the test would be independent of any specific assumptions as to the nature of the non-proton final states.
\vspace{2mm}

It would be truly amazing if the good old neutron turned out to be the particle enabling us to probe the dark  sector of the universe, binding even more closely the nuclear and particle physics communities. 
However, even if the neutron lifetime discrepancy gets resolved in favor of the Standard Model, the neutron dark decay with a reduced rate would still remain a viable scenario to consider in the quest of understanding the dark  side of our universe. 

\newpage

\bibliography{neutron}
\bibliographystyle{unsrt}

\end{document}

%% file: econfmacros.tex
\def\beq{\begin{equation}}
\def\eeq#1{\label{#1}\end{equation}}
\def\eeqn{\end{equation}}

\def\beqa{\begin{eqnarray}}
\def\eeqa#1{\label{#1}\end{eqnarray}}
\def\eeqan{\end{eqnarray}}

\let\bar=\overbar

\def\Dslash{\not{\hbox{\kern-4pt $D$}}}
\def\dslash{\not{\hbox{\kern-2pt $\del$}}}

\def\msb{{\bar{\ssstyle M \kern -1pt S}}}

